\documentclass[preprint2]{aastex}

\usepackage[english]{babel}
\usepackage[utf8]{inputenc}
\usepackage{amsmath}
\usepackage{graphicx}
\usepackage[colorinlistoftodos]{todonotes}
\usepackage{subfigure}
\usepackage{afterpage}
\usepackage{flushend}
\usepackage{indentfirst}
\usepackage{rotating}
\providecommand{\e}[1]{\ensuremath{\times 10^{#1}}}

\title{A Search for Fast X-ray Variability from Active Galactic Nuclei using \textit{Swift}}

\author{Matthew Pryal, Abe Falcone, Michael Stroh}
\affil{Department of Astronomy \& Astrophysics}
\affil{The Pennsylvania State University, University Park, PA 16802}

\date{}
\begin{abstract}
Blazars are a class of active galactic nuclei (AGNs) known for their very rapid variabilty in the high energy regions of the electromagnetic spectrum. Despite this known fast variability, X-ray observations have generally not revealed variability in blazars with rate doubling or halving timescales less than approximately 15 min. Since its launch, the \textit{Swift} X-ray Telescope has obtained 0.2-10 keV X-ray data on 143 AGNs, including blazars, through intense target of opportunity observations that can be analyzed in a multiwavelength context and used to model jet parameters, particularly during flare states. We have analyzed this broad \textit{Swift} data set in a search for short timescale variability in blazars that could limit the size of the emission region in the blazar jet. While we do find several low-significance possible flares with potential indications of rapid variability, we find no strong evidence for rapid ($<$15 minutes) doubling or halving times in flares in the soft X-ray energy band for the AGNs analyzed.   
\\

\end{abstract}

\keywords{acceleration of particles - galaxies: active - galaxies: jets - gamma rays: general - X-rays: general}

\begin{document}
\maketitle


\section{Introduction}
\label{sec:intro}

Analysis of the flux variability of active galactic nuclei (AGNs) allows constraints to be put on the associated emission mechanisms and jet parameters of the AGN jet. Flux variability in AGNs is seen across the entire electromagnetic spectrum with the fastest variability seen in the higher energy regions, in particular the keV to TeV regions. Of all AGNs, the strongest and fastest variability is often seen in a class of AGNs known as blazars, which are AGNs that have a jet pointed in a direction that is very close to our line of sight. Analysis of the flux variability of blazars in the high-energy region of the electromagnetic spectrum is a useful method for limiting the size of the emission region in an AGN, thereby limiting the emission mechanisms.

Due to light propagation time and geometry effects, fast variability can limit the size of the emission region R of a blazar by the inequality $R<(cT\delta)/(1+z)$ where T is the variability timescale of the flux, $\delta$ is the Doppler factor, and z is cosmological redshift. The Doppler factor $\delta$ is defined as $\delta^{-1}=\gamma(1-\beta cos\theta)$, where $\gamma$ is the Lorentz factor in the jet, $\beta$ is the ratio of the jet velocity to the speed of light in vacuum, and $\theta$ is the angle of the jet with respect to the observer. The variability timescale, T, is typically described by the flux doubling time, which will be referred to as T$_{double}$ or the flux halving time, which will be referred to as T$_{1/2}$.

There has been repeated evidence of fast flux variability in the TeV $\gamma$-ray region with doubling or halving times as short as 224$\pm$60 s from PKS 2155-304 (Aharonian et al., 2007), 15 minutes from Mrk 421 (Gaidos et al., 1996) and $\sim$ 9 min from BL Lac (Arlen et al., 2012). Results from simultaneous TeV and X-ray studies probing variability on the order of hours have shown evidence that X-rays and TeV $\gamma$-rays may originate from the same region (e.g. Maraschi et al. 1999). This correlation at larger timescales, in addition to observed fast variability of TeV $\gamma$-rays, has motivated the search for fast variability in blazars at X-ray wavelengths.

In general, X-ray observations have not revealed variability with doubling timescales less than approximately 15 min. There has been isolated evidence for much faster variability in X-rays from a single blazar, H0323+022 (Feigelson et al., 1986), with a halving time less than 30 seconds and a quoted statistical significance of P $<10^{-13}$. However, this fast variability in X-rays was a single, isolated event that has not been seen again. The authors did consider systematic and/or instrumental effects, but none could be identified.


With the launch of the \textit{Swift} X-Ray Telescope (Burrows et al., 2005), X-ray data in the 0.2-10 keV region has been obtained on many AGNs on timescales ranging from seconds to over 8 years.

We report on our search for fast variability in blazar flares. We have searched our large \textit{Swift} database for blazar flares with doubling or halving times less than 15 minutes. The data taken from the \textit{Swift} database for analysis are outlined in Section \ref{sec:data}. The analysis techniques, including the necessary constraints set on the data in the search for fast blazar variability, is outlined in Section \ref{sec:analysis}. We provide the results of the analysis in Section \ref{sec:results} with discussion of the results in Section \ref{sec:discussion}.

\section{Data}
\label{sec:data}

These data were obtained with the X-Ray Telescope (XRT) on the \textit{Swift} observatory (Burrows et al. 2005), and the reduced light curves are all available on a public web site and database described by Stroh \& Falcone (2013). The \textit{Swift}-XRT is one of three instruments onboard the \textit{Swift} Gamma Ray Burst Explorer (Gehrels et al., 2004).  In particular, the \textit{Swift}-XRT is an X-ray imaging spectrometer with the ability to obtain data for lightcurves with high timing resolution. The Swift-XRT data in this study are from two different modes, window timing mode (WT) and photon counting mode (PC).  All data are screened to ensure that no individual observations, and therefore no potential flares, include a switch from one mode to the other. See Burrows et al., 2005 for an in depth discussion of the XRT modes.

The data are comprised of 7.43\e{6} seconds of AGN observations from 12544 continuous observations of the 143 AGN observed at the time of analysis. These data were obtained between 2004 December 17 and 2014 May 28. The data were processed using the most up to date version of \textit{Swift} tools at the time of analysis: \textit{Swift} Software version 4.1 and FTOOLS version 6.14 (Blackburn, 1995) with observations processed using xrtpipeline version 0.12.8. For WT mode data, we additionally filtered out the first 150 seconds of data after a \textit{Swift} telescope slew since the telescope pointing was not always settled well enough to allow accurate standard WT mode data analysis. Greater detail on how these data were obtained and processed is available in Stroh \& Falcone (2013). 

The number of observations for each object range from as many as 1192 observations in the case of Mrk 421 to as little as 1 observation in the case of PKS 0244-470. The average continuous observation time is about 592 s, with a maximum observation time of 2565 s and a minimum observation time of 21 s. 

\afterpage{%
\begin{figure*}[ht!]
     \begin{center}
        \subfigure{%
            \label{fig:1st}
            \includegraphics[width=0.3\textwidth]{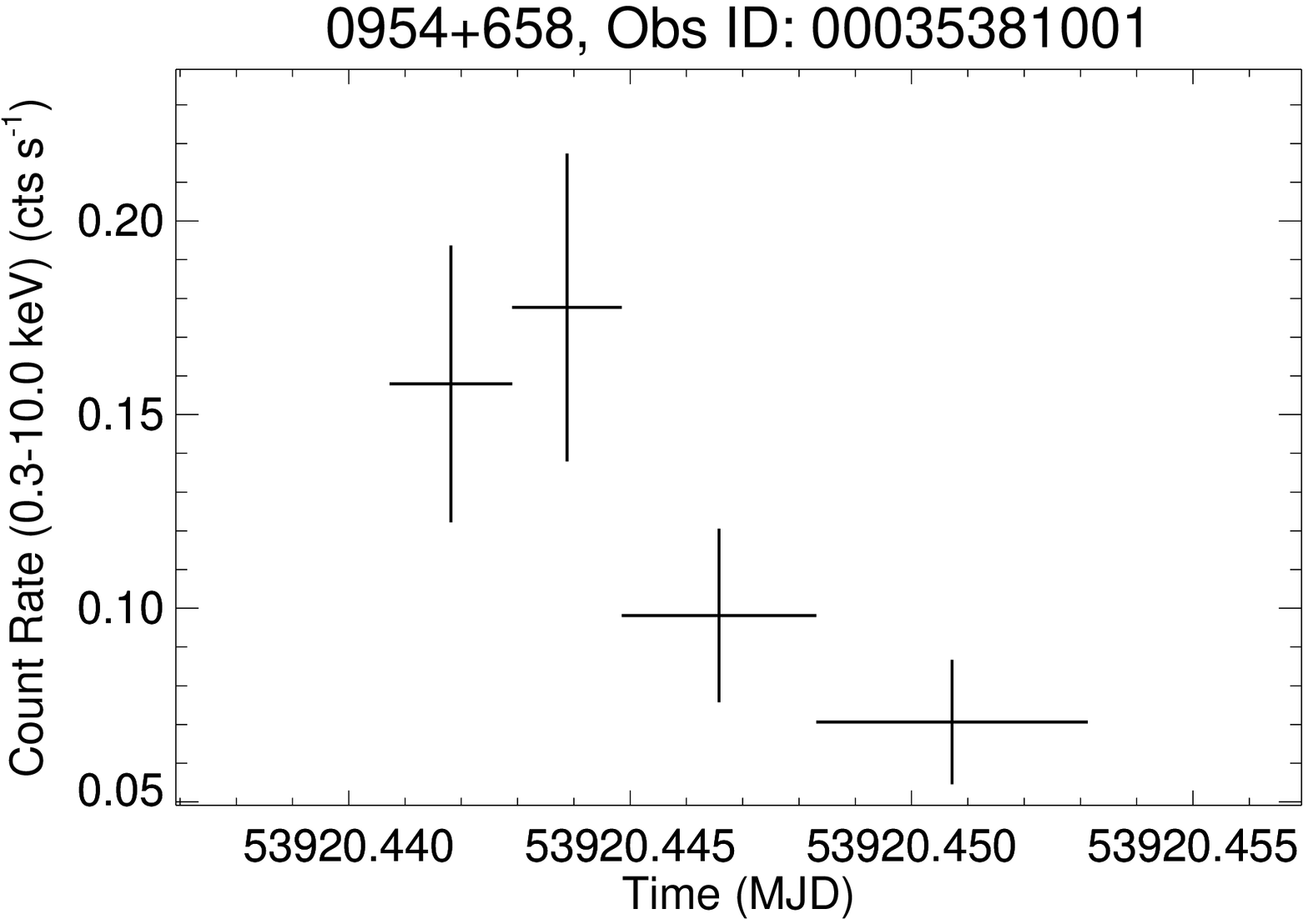}
        }%
        \subfigure{%
            \label{fig:2nd}
            \includegraphics[width=0.3\textwidth]{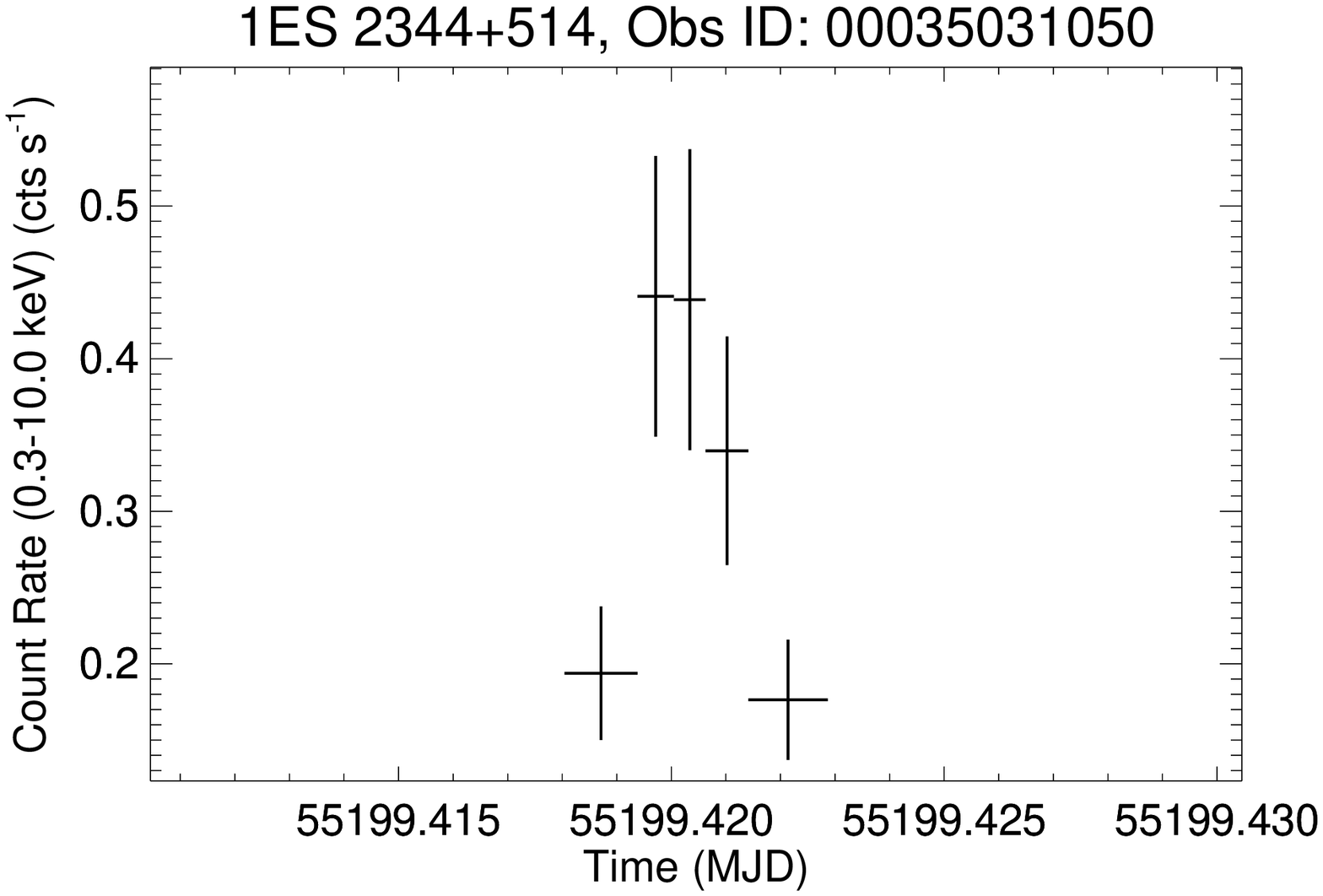}
        }%
        \subfigure{%
            \label{fig:3rd}
            \includegraphics[width=0.3\textwidth]{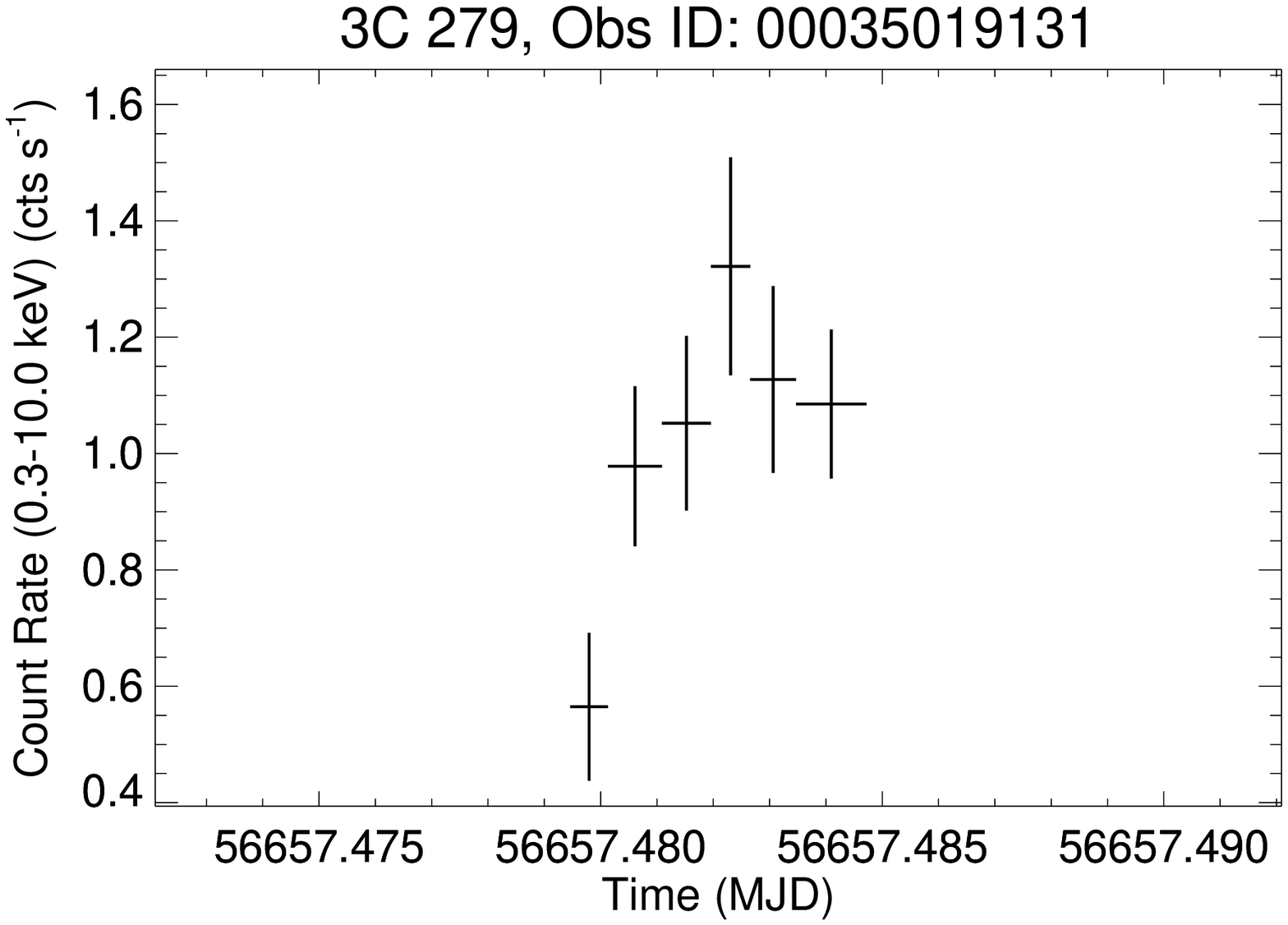}
        } \\
        \subfigure{%
            \label{fig:4th}
            \includegraphics[width=0.3\textwidth]{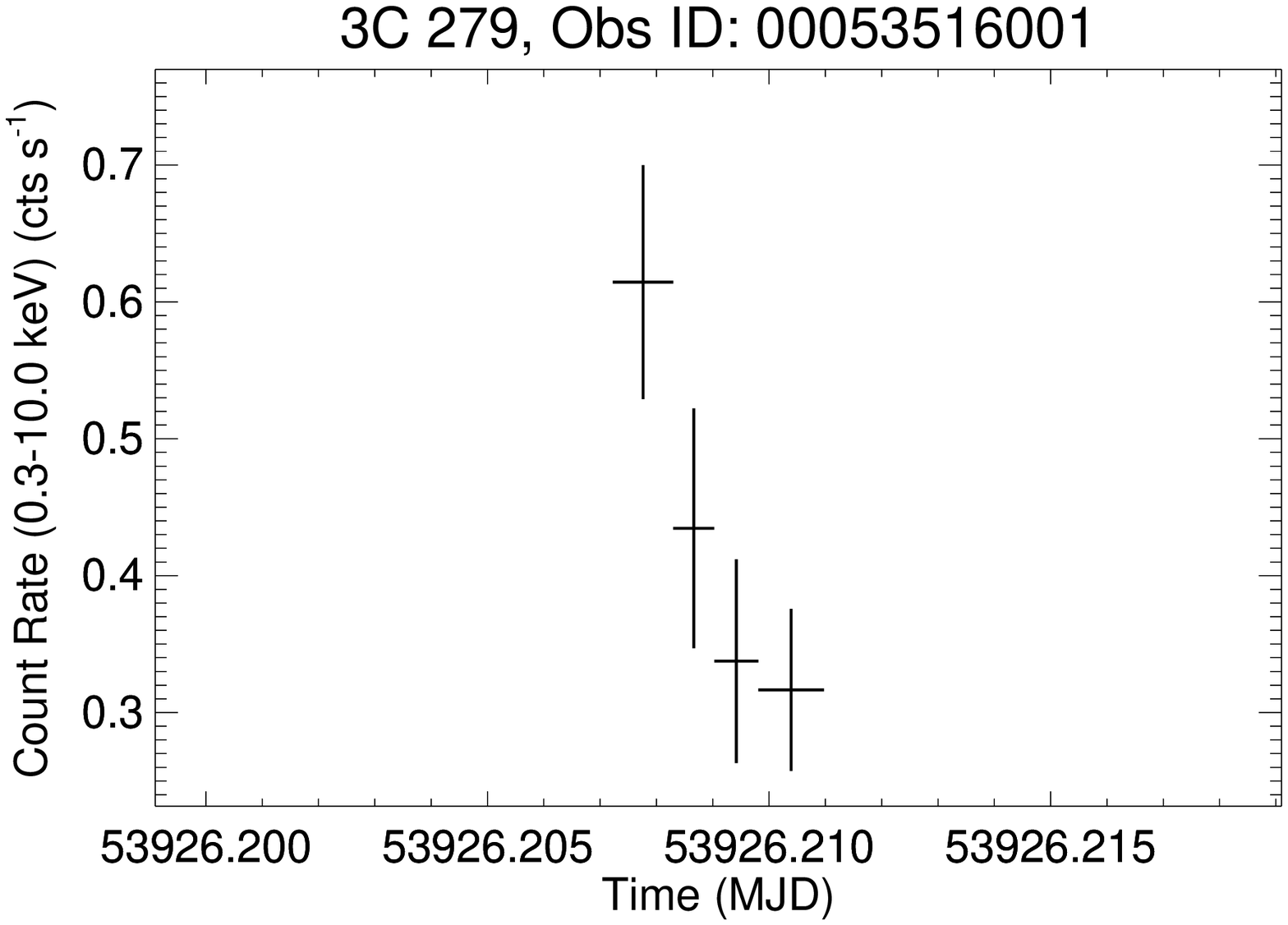}
        }       
        \subfigure{%
            \label{fig:5th}
            \includegraphics[width=0.3\textwidth]{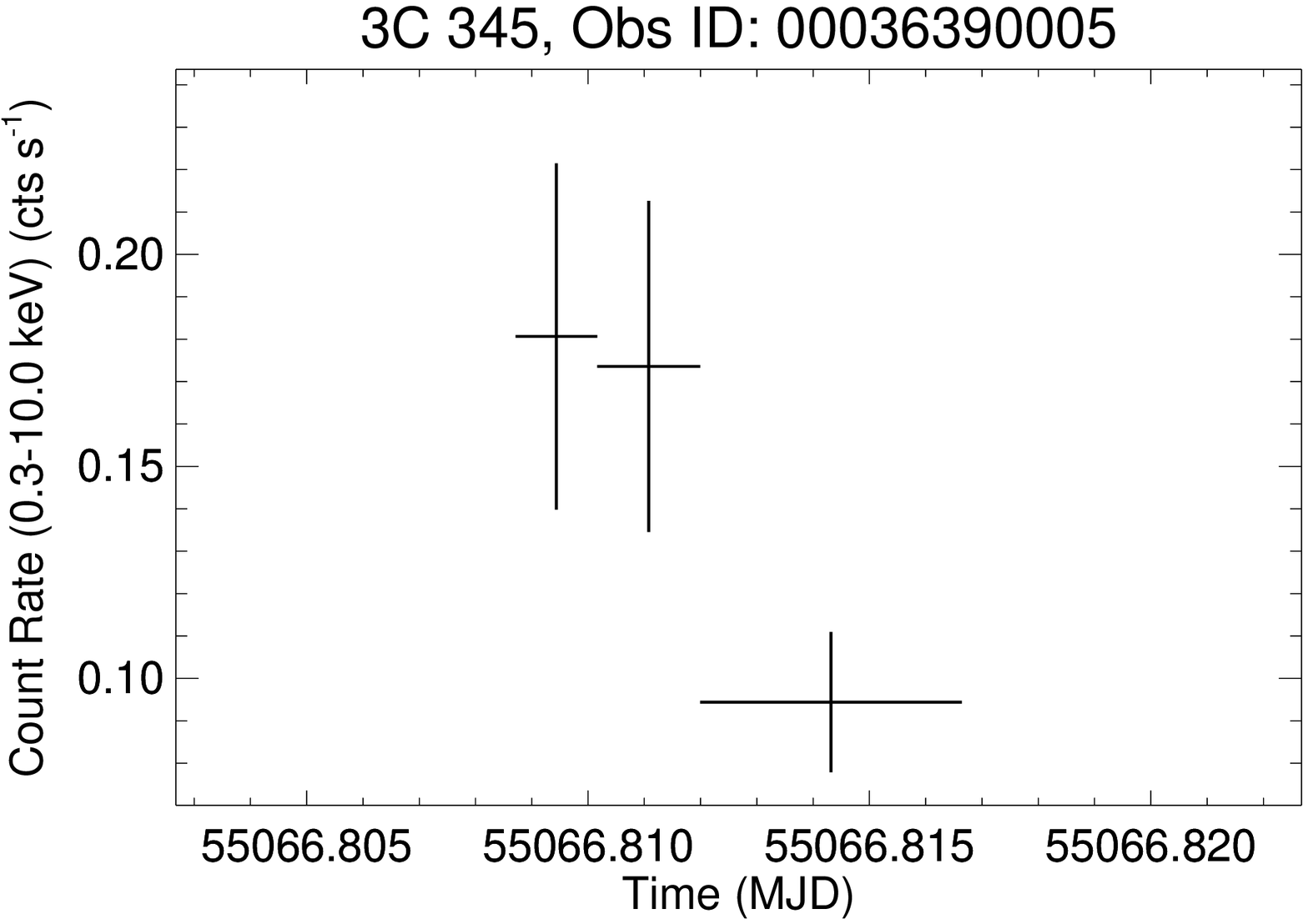}
        }%
        \subfigure{%
            \label{fig:6th}
            \includegraphics[width=0.3\textwidth]{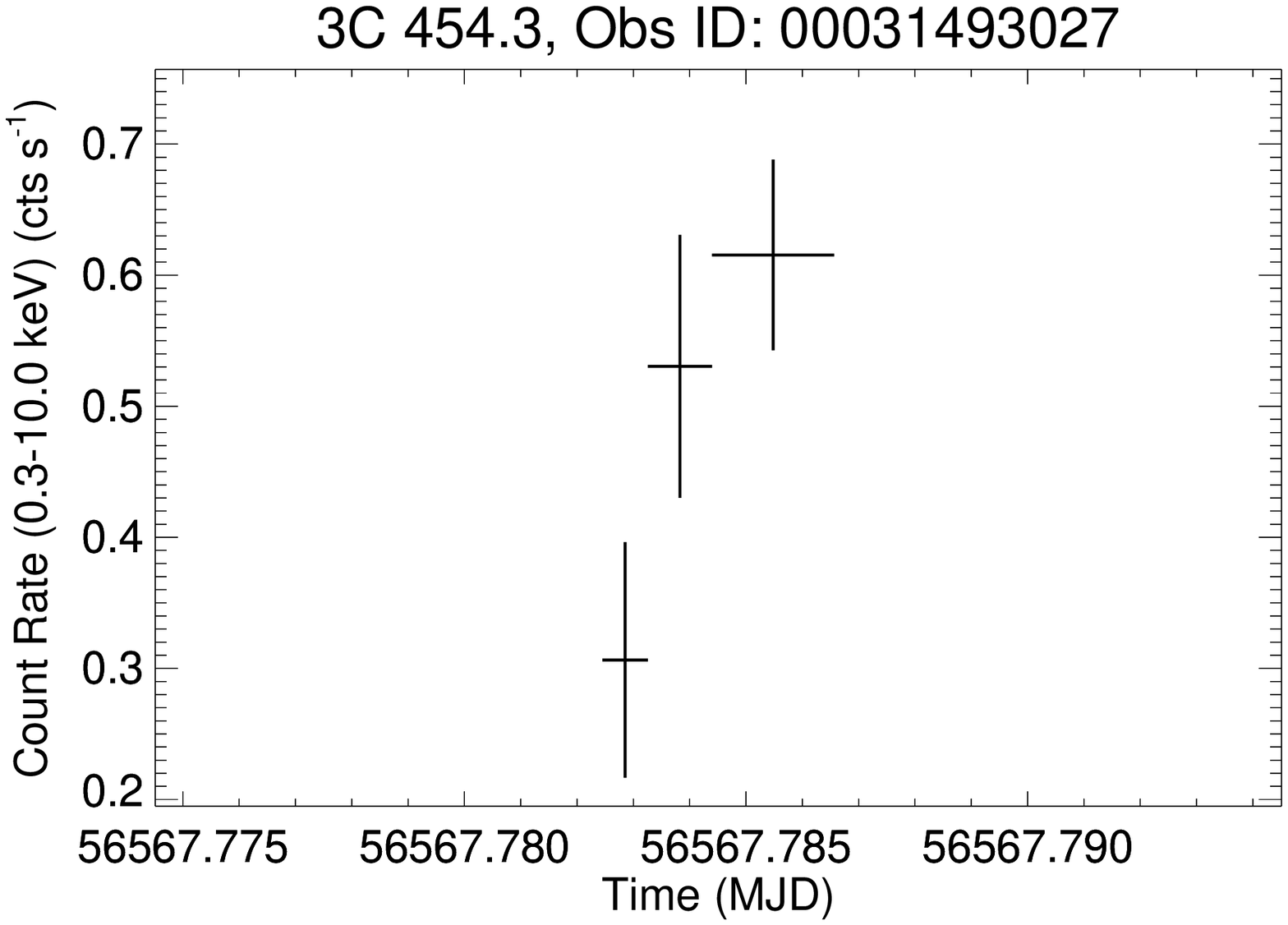}
        } \\
        \subfigure{%
            \label{fig:7th}
            \includegraphics[width=0.3\textwidth]{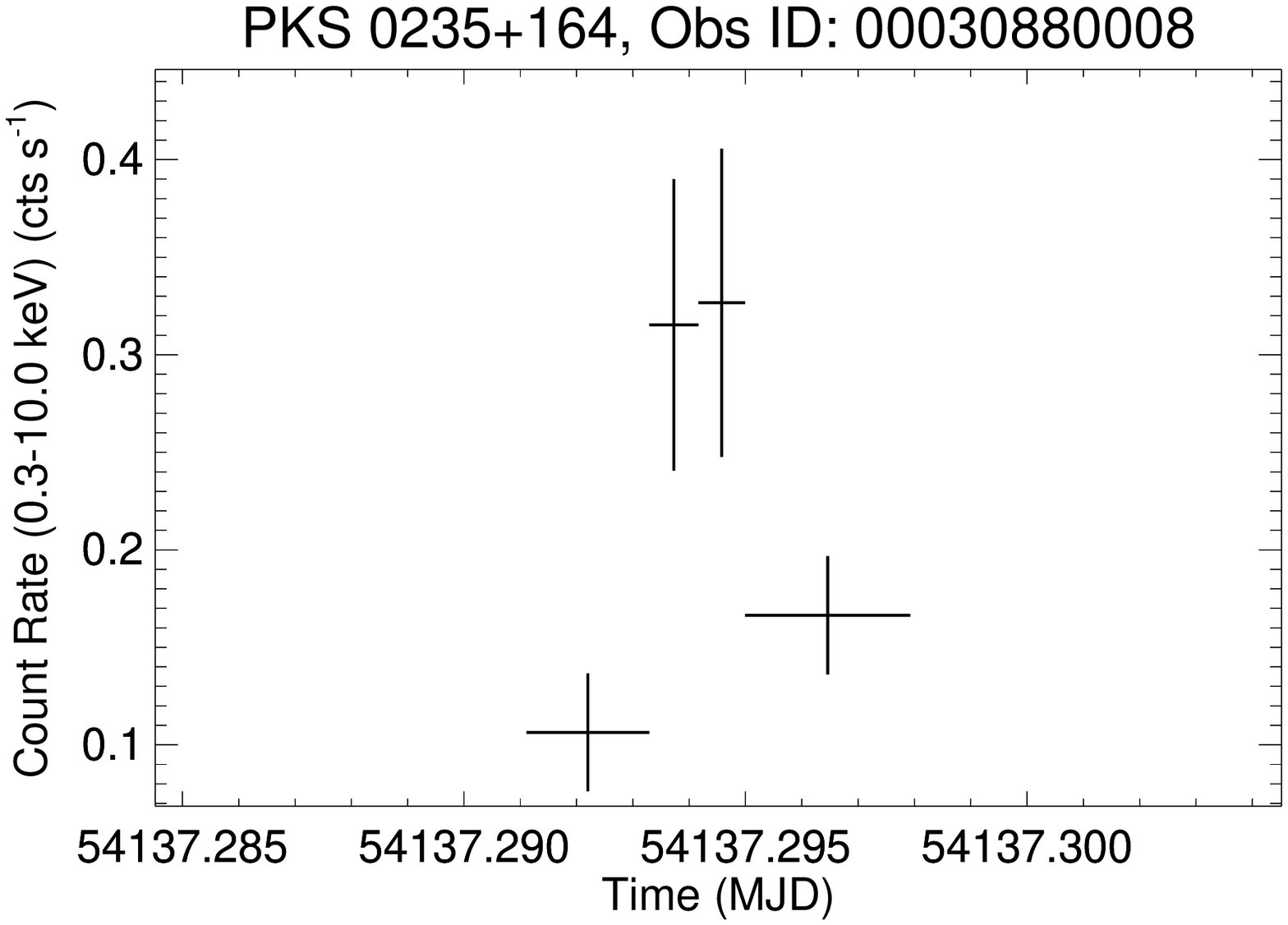}
        }
        \subfigure{%
            \label{fig:8th}
            \includegraphics[width=0.3\textwidth]{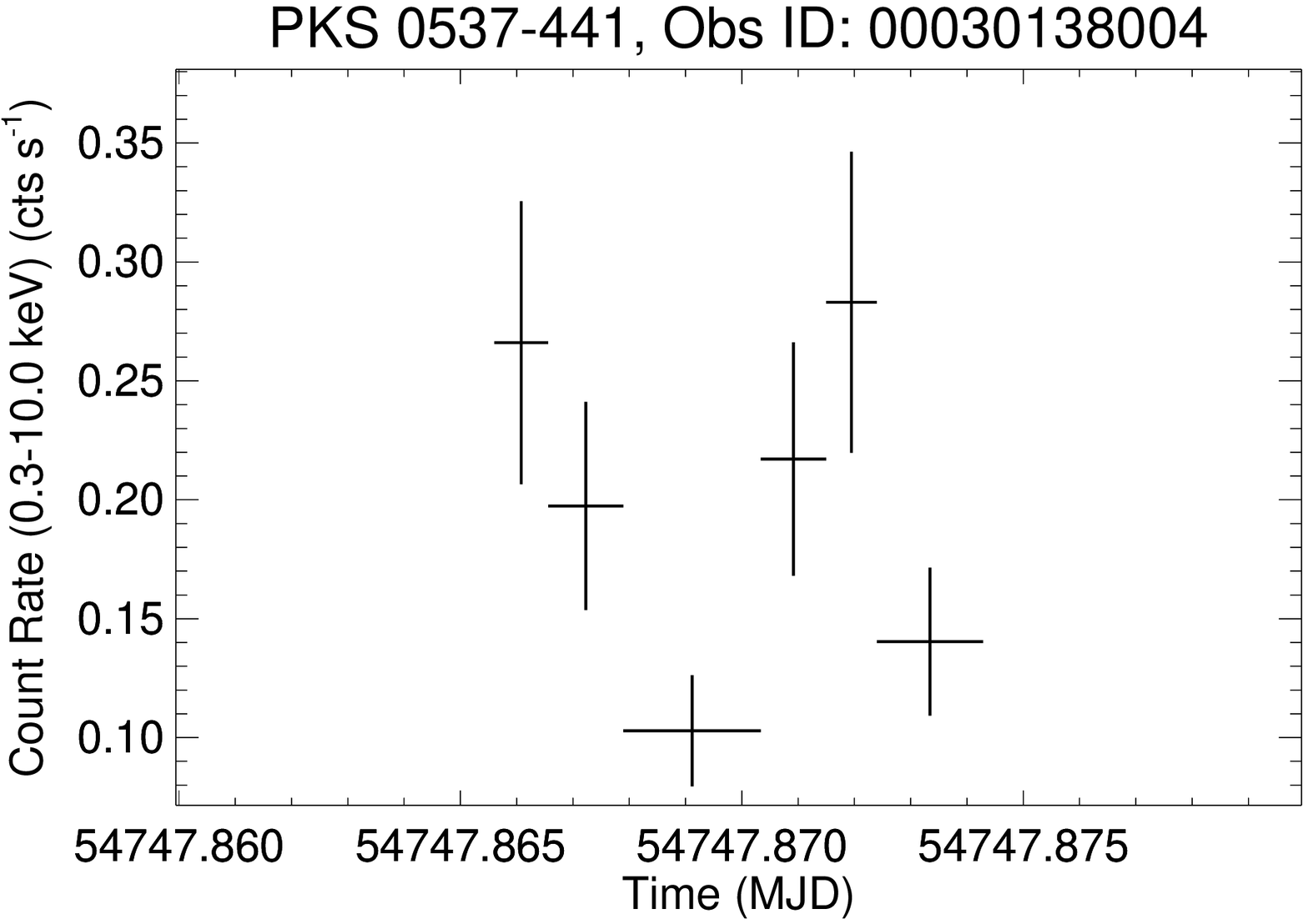}
        }%
        \subfigure{%
            \label{fig:9th}
            \includegraphics[width=0.3\textwidth]{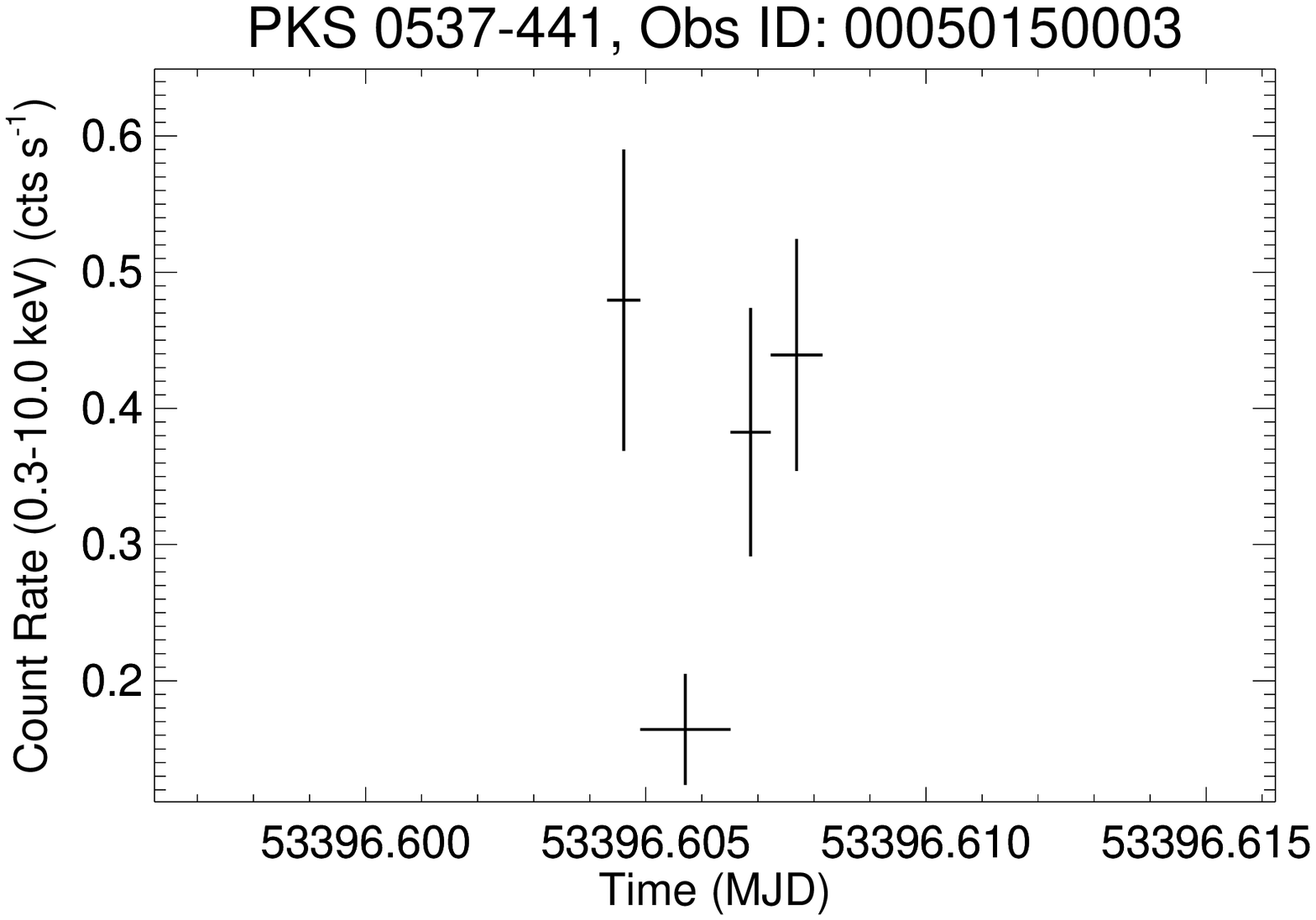}
        }\\%
        \subfigure{%
           \label{fig:10th}
           \includegraphics[width=0.3\textwidth]{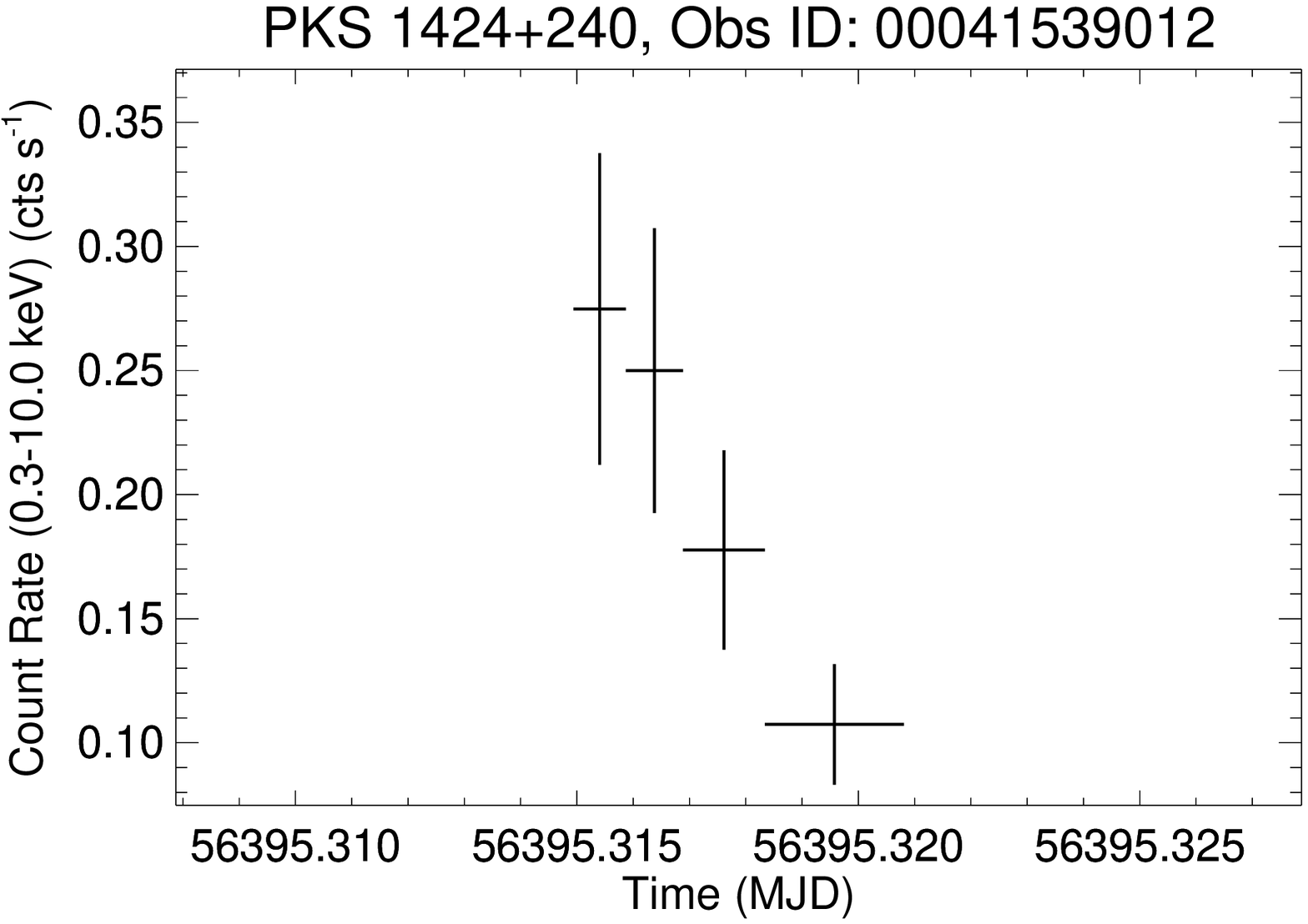}
        } 
        \subfigure{%
            \label{fig:11th}
            \includegraphics[width=0.3\textwidth]{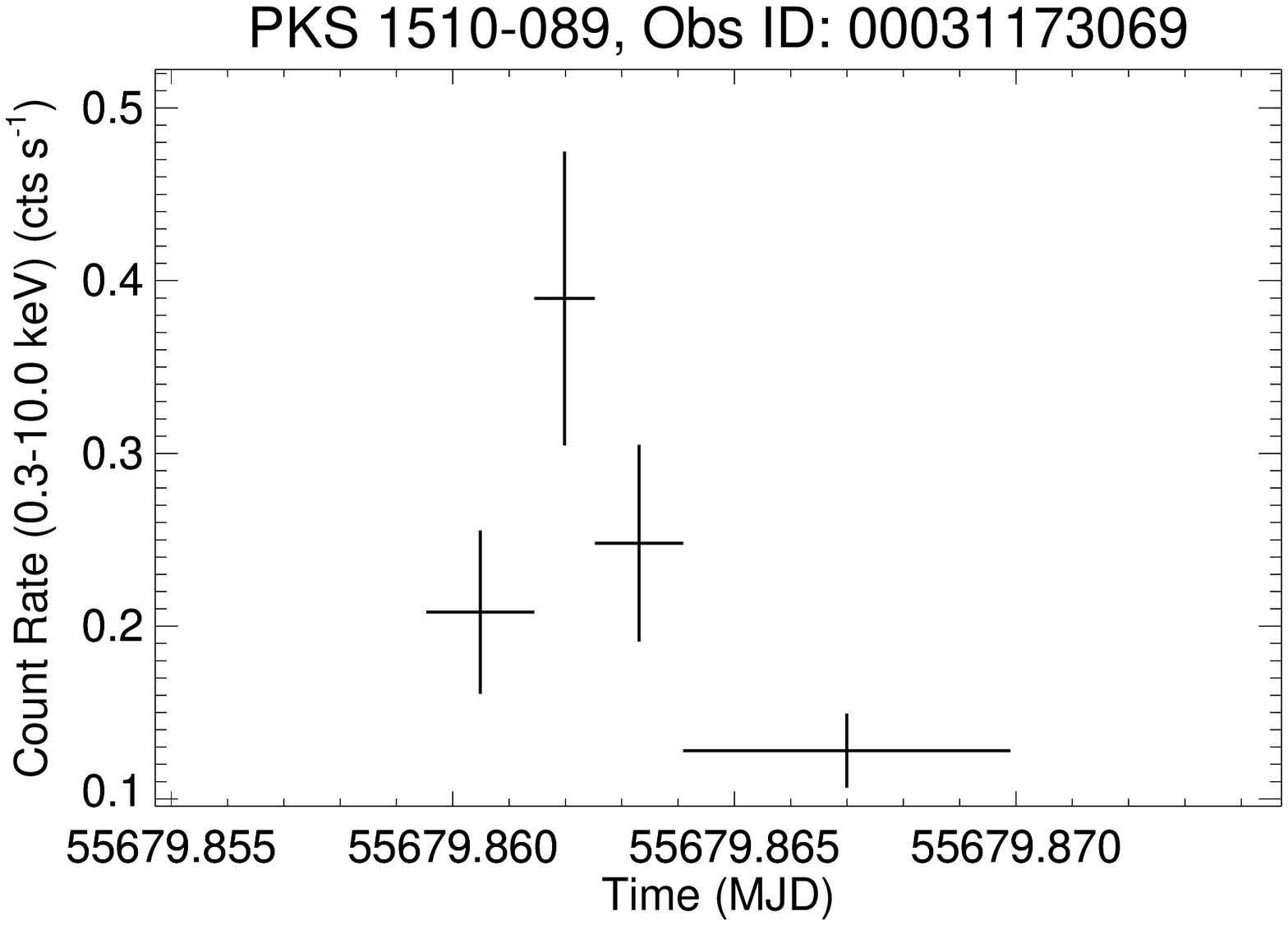}
        }
        \subfigure{%
            \label{fig:12th}
            \includegraphics[width=0.3\textwidth]{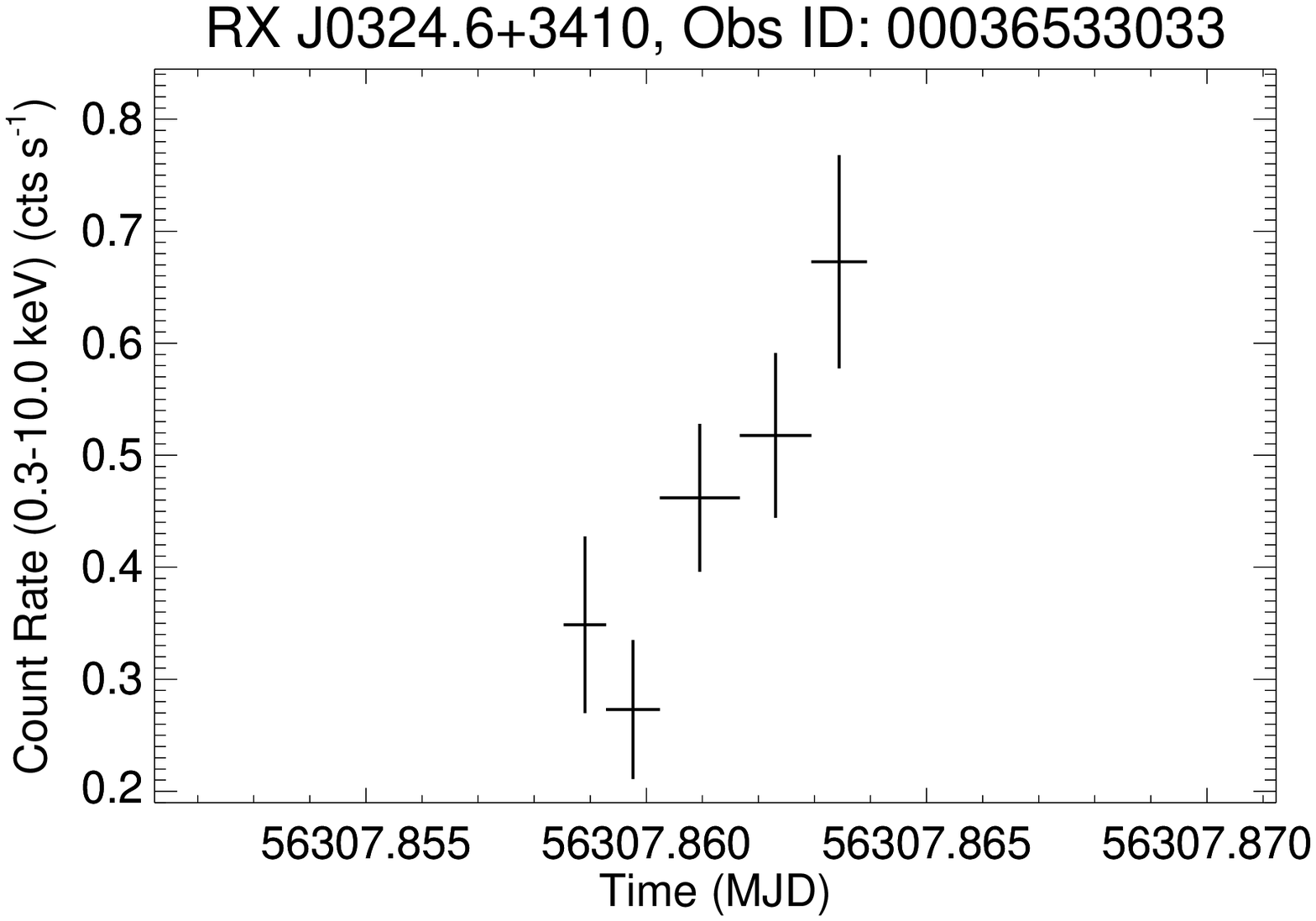}
        }\\%
        \subfigure{%
           \label{fig:13th}
           \includegraphics[width=0.3\textwidth]{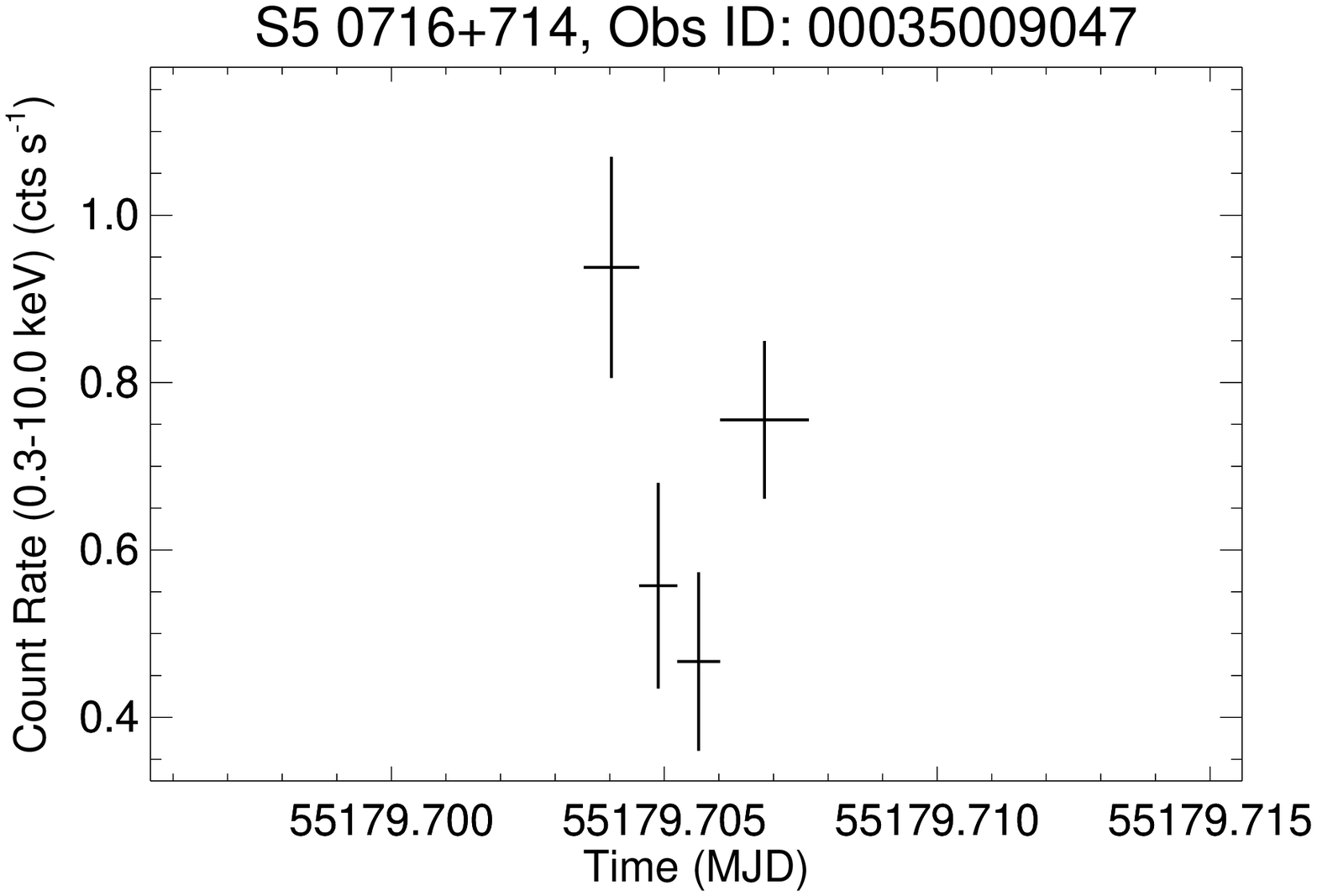}
        } 
        \subfigure{%
            \label{fig:14th}
            \includegraphics[width=0.3\textwidth]{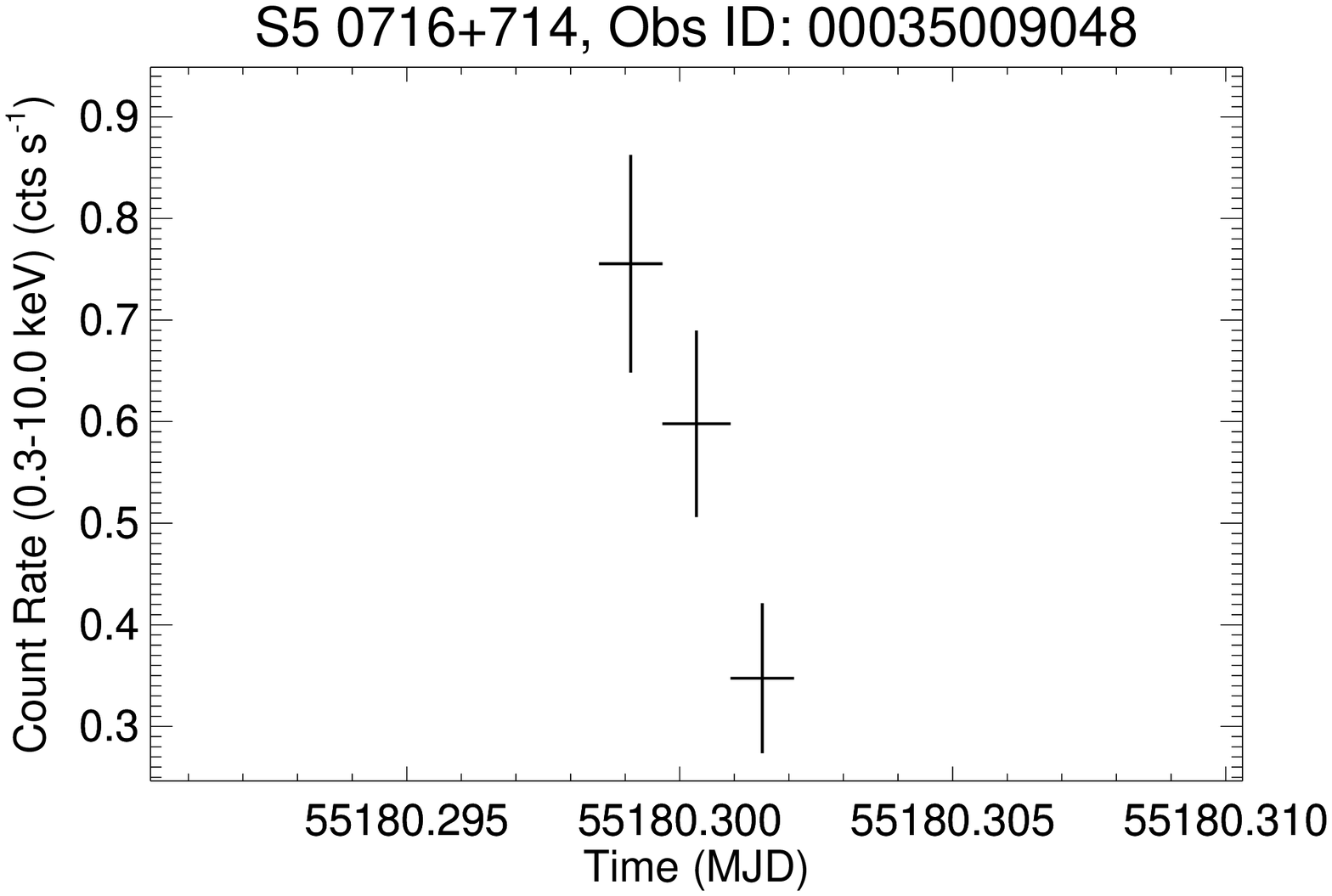}
        }%
        \subfigure{%
           \label{fig:15th}
           \includegraphics[width=0.3\textwidth]{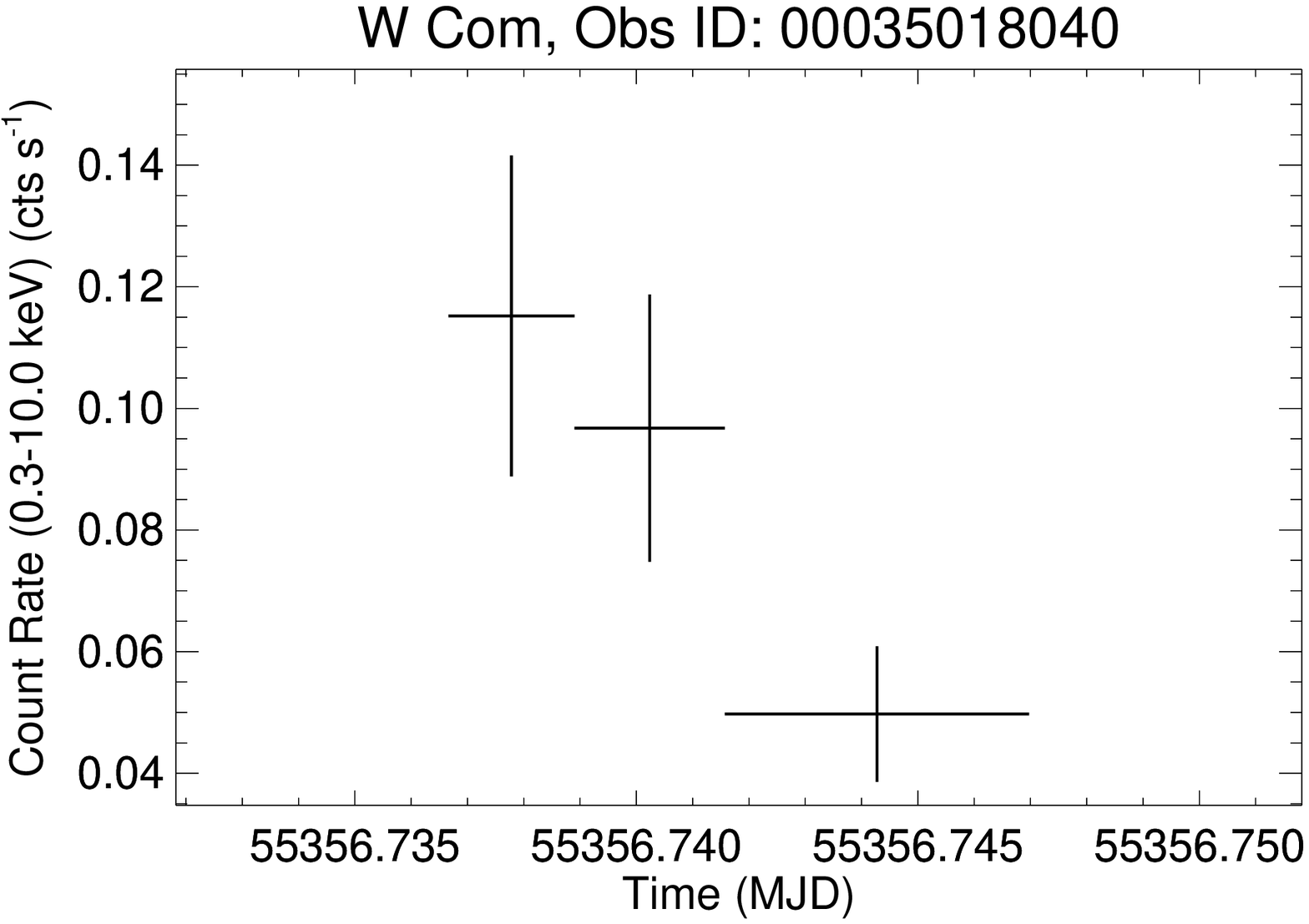}
        }
    \end{center}
    \caption{%
        The light curves of the 15 observations flagged for a potential quick time X-ray flare are shown. The observation ID specified in the title of each plot is the observation ID assigned by \textit{Swift}.
     }%
   \label{fig:plots}
\end{figure*}
\clearpage
}

\section{Analysis}
\label{sec:analysis}

The goal of the analysis is to determine if fast blazar variability with a rate doubling or halving time less than 15 minutes exists within these data. 
The data were analyzed on an observation by observation basis to test for this fast variability. Various constraints were needed to: (1) define the data set of observations that would have enough counts and a long enough duration to provide the potential to find significant flaring, and (2) define the observations within the aforementioned data set that would have variability, and in particular, the potential for flux doubling or halving. The necessary constraints to define our data set include: a maximum count rate of at least 0.1 cts/s, at least three data points in each observation, and a length of observation of at least 5 minutes. The constraints that define a possible flare include: restricting the maximum count rate to be at least 1.8 times greater than the minimum count rate during the observation, 3 consecutive data points either increasing or decreasing in count rate, and a reduced $\chi^2$ value from a straight line fit greater than 3.0. In addition to these pre-defined constraints, manual inspection was required for the observations that passed the constraints to ensure that flux variations from systematic effects (e.g. mode switching) were not flagged as possible flares. The necessity of each of these constraints is discussed in more detail below.

\subsection{Constraints That Define Data Set}
\label{subsec:definedataset}

\subsubsection{Max Count Rate}

A maximum count rate of at least 0.1 cts/s during an observation is necessary to allow for a statistically significant flare to occur in the timescales being probed. Taking into account light curve error bars at this count rate and applying $\chi^2$ statistics, a statistically significant flare cannot be detected at rates lower than 0.1 cts/s over the timescales of concern. Therefore, if the maximum count rate of an observation is not at least 0.1 cts/s then it would be impossible to detect a fast flare from that observation, thus we eliminate such observations from the data sample.

\subsubsection{Data Point Requirement}

The requirement of at least three data points within an observation as a means of defining our data set is closely related to the discussion in Section \ref{subsub:consecutive}. We cannot include observations that have less than three data points as part of our data set because they do not have the ability to have three increasing or decreasing data points and therefore would not possibly be able to be flagged as a possible flare.

\subsubsection{Orbit Length}

We also require observations to be at least 5 minutes in length. This is to allow enough time for a significant flare to occur within a typical observation. Our data set also has max observation time of 2565 seconds, which implicitly sets an upper limit to the length of the doubling or halving time observed of a potential flare, though no explicit cut was set.

\subsection{Constraints That Define Flaring}
\label{subsec:defineflare}

\subsubsection{Doubling/Halving Factor}

We require the maximum count rate to be 1.8 times greater than the minimum count rate to ensure that any potential flares at least approach a flux doubling or halving since the purpose of the analysis is to search for AGN flares with fast doubling or halving timescales. 

\subsubsection{Consecutive Points}
\label{subsub:consecutive}

We require at least 3 consecutive data points to be either increasing or decreasing in count rate to minimize random noise in the count rate and guarantee a possible steady increase or decrease in the flare. 

\subsubsection{Reduced $\chi^2$ Value}
\label{subsub:chi2}

Finally, a reduced $\chi^2$ value from a straight line fit of the weighted mean of the count rate greater than 3.0 is necessary to ensure sufficient variability in the count rate of the source. This constraint ensures that the selected observation is not consistent with a quiescent state of the AGN. This reduced $\chi^2$ value was also used in determining the p-value of the null hypothesis of a straight line fit. 

\begin{figure}[ht!]
     \begin{center}
        \subfigure{%
            \label{fig:afirst}
            \includegraphics[width=0.4\textwidth]{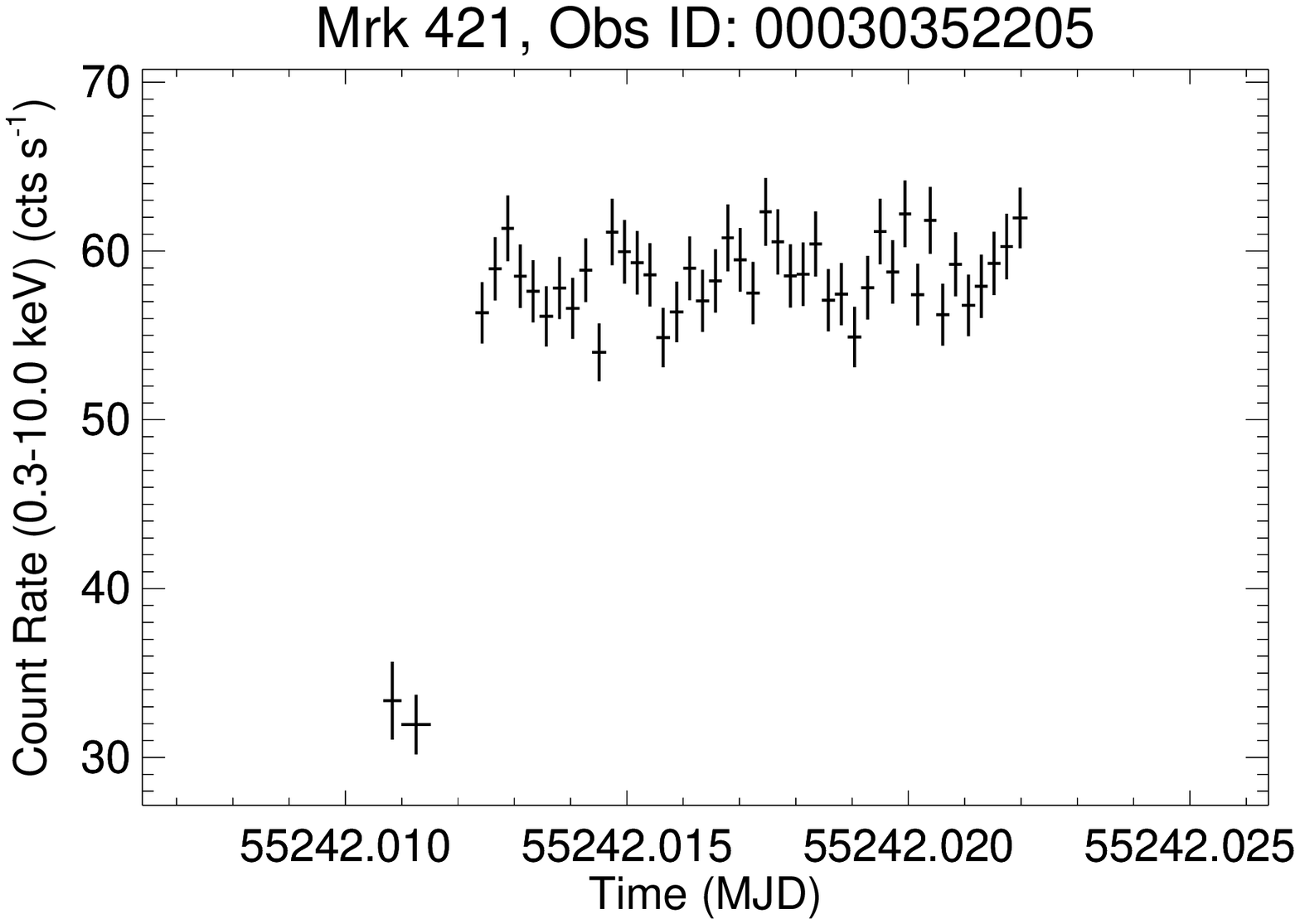}
        } \\
        \subfigure{%
           \label{fig:asecond}
           \includegraphics[width=0.4\textwidth]{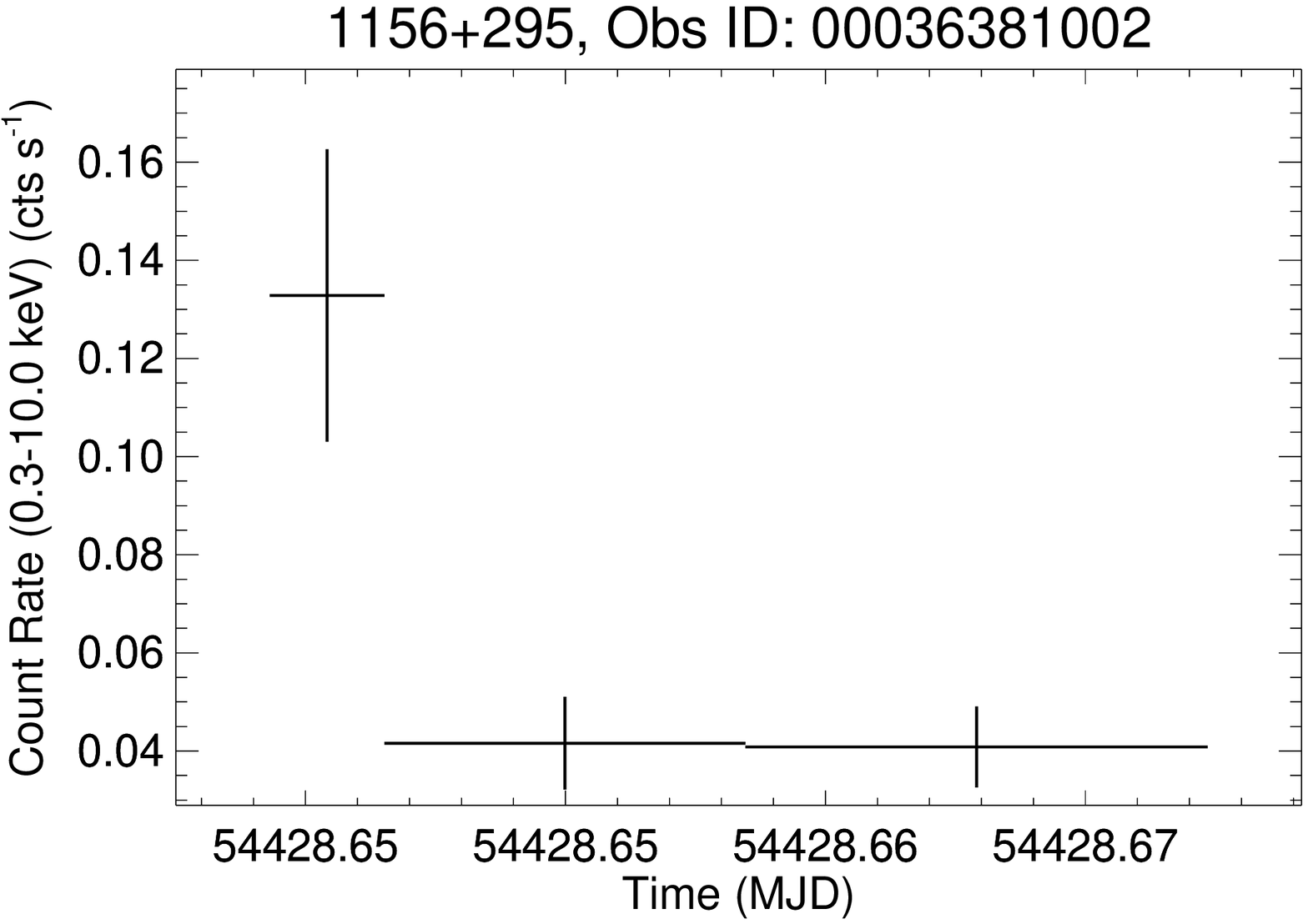}
        } 
    \end{center}
    \caption{%
        The light curve from Mrk 421 (top) shows a sharp break in the light curve that results in spurious data points as the result of the \textit{Swift}-XRT switching observing modes from PC mode to WT mode. The light curve of 1156+295 (bottom) shows a single spurious data point that was potentially a result of fluctuations at the beginning of an observation that could not be considered a potential quick time flare.
     }%
   \label{fig:spurious}
\end{figure}

\subsection{Manual Analysis}
\label{subsec:manual}

Manual inspection was required to ensure that all observations flagged as possible flares did not have any issues that could not be flagged systematically. One of these issues includes the \textit{Swift}-XRT mode switching from PC to WT or vice versa, which often caused an abrupt break in the light curve and spurious points to occur in the data as shown in light curve on the top of Figure \ref{fig:spurious}. This issue would occasionally cause observations to be flagged as a possible flare and would have to be discarded manually. Other observations that were processed out through manual inspection had a single spurious data point responsible for the identification of the observation as a potential flare, with the single point occurring either at the beginning or end of an observation with no flaring evidence throughout the rest of the light curve. These observations (e.g. Figure \ref{fig:spurious}, bottom) also had to be discarded from the data set.

\section{Results}
\label{sec:results}

Of the initial 12544 AGN observations in the large data set, 8606 AGNs observations had the possibility of having a detectable quick time flare based on the count-rate constraints and the a priori assumptions that define our data set as discussed in Section \ref{subsec:definedataset}. Of the remaining 8606 AGN observations, 31 observations passed the constraints that defined a possible flare as discussed in Section \ref{subsec:defineflare}. Once the manual analysis was done, 21 observations remained that did not have any mode switching issues or were reliant on a single spurious data point, as discussed in Section \ref{subsec:manual}. Of the remaining 21 observations, six were flagged as being clearly due to random fluctuations, without any clearly discernible flare, and were therefore not considered to be potential flares, and one observation from PKS 0537-441 showed signs of two flaring episodes within a single observation. Therefore, 16 potential flares from 15 observations and 12 different AGN remained. Details from these 16 potential flares are shown in Table \ref{table:orbits}, and each light curve is shown in the mosaic of Figure \ref{fig:plots}.

We used two methods to estimate the significance of the quick time soft X-ray flares; one method compares the light curve data to an assumption that the X-ray rate was constant with a value equal to the mean rate during the observation, and a second method compares the light curve data to an assumption that the X-ray rate was constant with a value equal to a best guess at the quiescent source rate near the time of the potential flare. In the first method, the chance probability p-value was determined based on the null hypothesis probability of a straight line fit of the weighted mean and using the reduced $\chi^2$ value that was found as discussed in Section \ref{subsub:chi2}. In the second method the chance probability p-value was determined based on the null hypothesis probability of the light curve data being described by a constant, straight-line fit to the minimum rate surrounding each potential flare, which was used as a best-guess estimate of the "quiescent" state of emission surrounding the potential flare. We computed the reduced $\chi^2$ value of the data during the potential flare relative to this "quiescent" constant rate light curve and then determined the p-value. These values are shown in Table \ref{table:orbits} under "Weighted Mean p-value" and "Quiescent p-value" respectively.

Once the p-value was determined using the two different methods, we had to take into account a trials factor based on the number of observations in the data set that had the potential to provide a detection of a rapid soft X-ray flare. The number for the trials factor was determined by counting the number of observations that passed the set constraints as discussed in Section \ref{subsec:defineflare} and Section \ref{subsec:manual}, as well as the visual inspection described above. Therefore, there were 16 trials that were taken into account, and the resultant p-values for each method are shown in Table \ref{table:orbits} in the "Post-trials" column next to each respective method.

\begin{figure}[ht!]
     \begin{center}
        \subfigure{%
            \label{fig:bfirst}
            \includegraphics[width=0.4\textwidth]{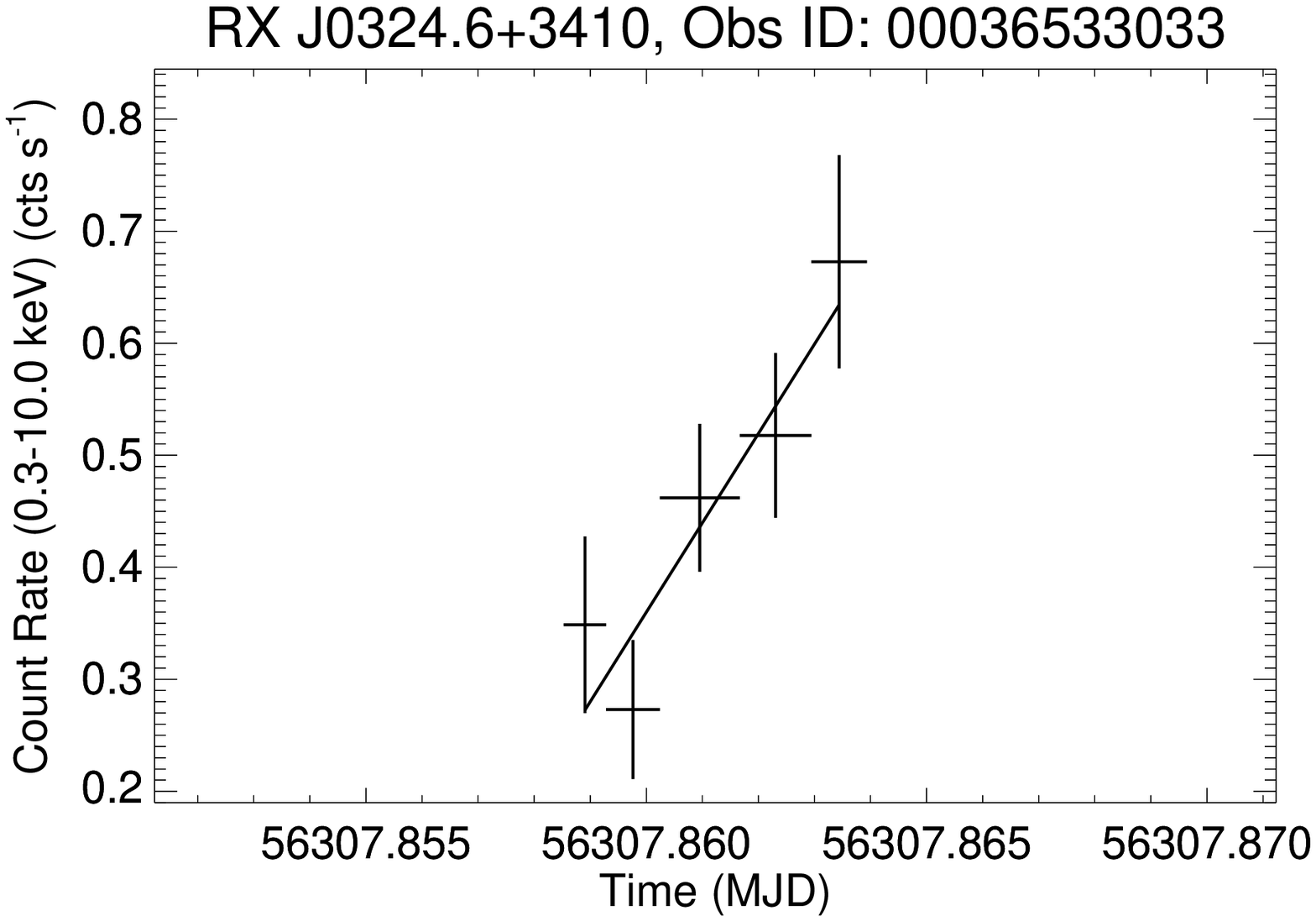}
        } \\
        \subfigure{%
           \label{fig:bsecond}
           \includegraphics[width=0.4\textwidth]{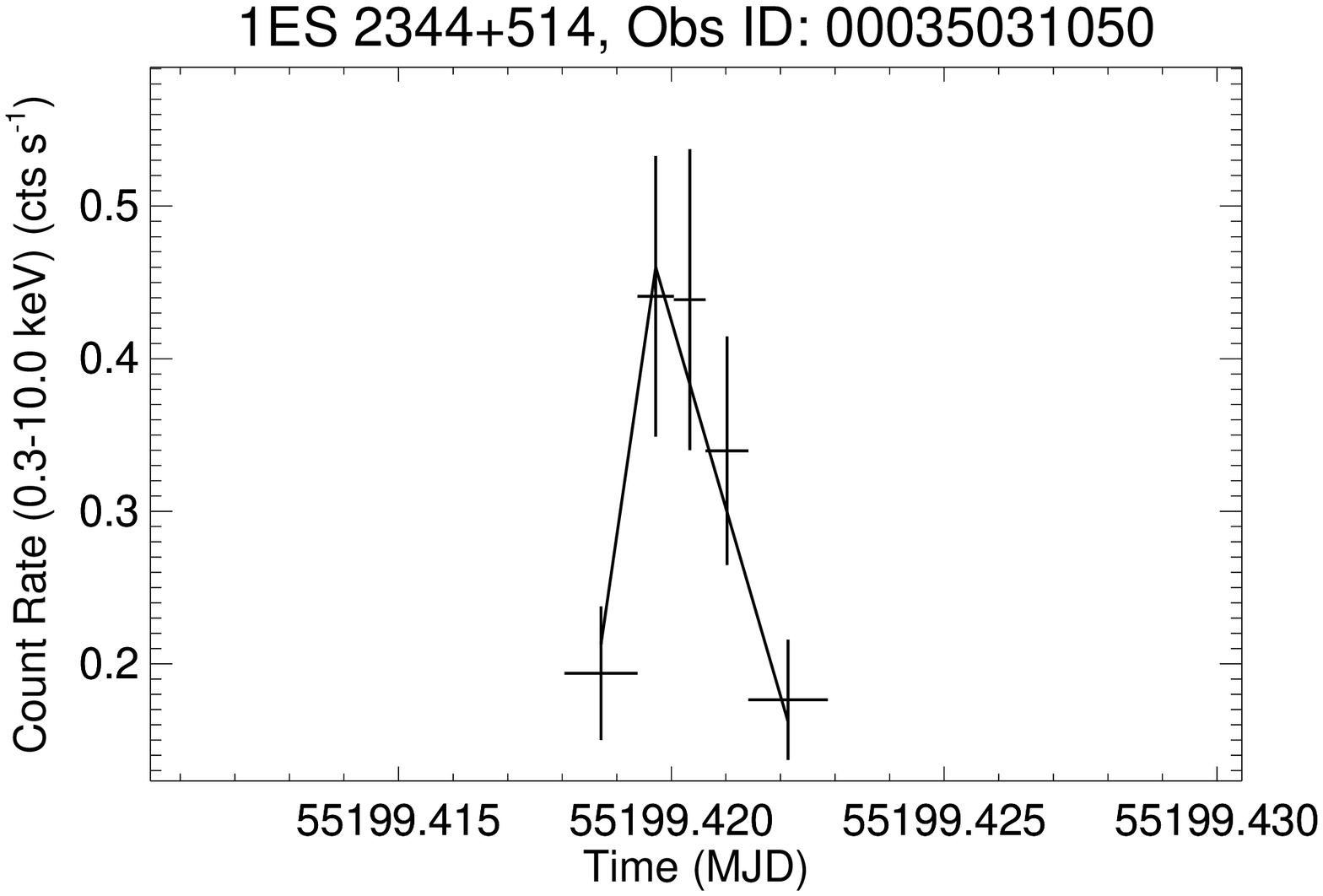}
        } 
    \end{center}
\caption{\label{fig:double}These plots show the selected section of the observations from RX J0324.6+3410 on 2013 Jan 15 (top) and 1ES 2344+514 on 2012 January 3 (bottom) that were fit with either increasing or decreasing lines to estimate the doubling or halving time.}
\label{fig:RXJ0324}
\end{figure}

In Table \ref{table:orbits} we also include the reduced $\chi^2$ values that were found on the section of each observation that was deemed part of the potential quick time flare as discussed earlier for the weighted mean and quiescent straight line fits. We also include the doubling times and halving times when applicable to each flagged observation. These times were determined by fitting a straight line to the sections of each observation deemed to be part of a possible flare and calculating how much time would be required for half of the maximum count rate to double or for the maximum count rate to halve when appropriate. An example of this fitting can be seen in Figure \ref{fig:double}. We also include the date of observation of each potential flare as well as the start time of the observation containing the possible flare as a reference to compare the data to the light curves available on the \textit{Swift}-XRT monitoring site (Stroh \& Falcone, 2013).

\section{Discussion}
\label{sec:discussion}

Analysis of our large \textit{Swift}-XRT database indicates that our data set does not show strongly significant evidence for soft X-ray flaring on the timescales probed, i.e. less than 15 minutes. Of the low-significance potential flares that were found and outlined in Table \ref{table:orbits} and Figure \ref{fig:plots}, we find potential doubling times in 6 different AGNs between 2.2 min and 5.7 min as well as halving times in 10 different AGNs between 2.2 min and 9.1 min. All of the potential flares in Table \ref{table:orbits} come from the blazar class (FSRQs or BL Lacs), which are more likely to exhibit rapid flaring due to their aligned relativistic jets. However, none of these potential flares represents a truly significant detection after accounting for trials, and therefore, no strong statements can be made about the existence, or nonexistence, of short timescale flares in the data set.

The potential flare from the RX J0324.6+3410 observation is shown in more detail in Figure \ref{fig:RXJ0324} (top). This potential flare has a doubling time, calculated as discussed in Section \ref{sec:results}, of 5.7 minutes. If the variability is truly from flaring, the potential emission region size can be calculated by applying the inequality $R<(cT\delta)/(1+z)$, which was discussed in Section \ref{sec:intro}. The doubling time variabilitiy as well as z=0.061 (Linford et al. 2012) equate to an upper limit on the emission region size of 9.7\e{12}$\delta$ cm, where $\delta$ is the unknown doppler factor that was discussed in Section \ref{sec:intro}.  

If the potential flares outlined in Table \ref{table:orbits} were indeed quick time soft X-ray flares, these would be the fastest doubling and halving times for X-ray AGN flares ever observed, with the exception of the single, isolated event observed by Feigelson et al. (1986). As stated previously, variability in AGNs at timescales approximately this short had previously only been seen repeatedly with statistical significance in the TeV region (Aharonian et al., 2007; Gaidos et al., 1996, Albert et al., 2007).

Simultaneous multiwavelength studies of blazars in TeV $\gamma$-rays and X-rays are vital to determine if very fast X-ray variability is coincident with very fast TeV variaiblity as has been seen previously on longer timescales (Maraschi et al., 1999). Further X-ray observations of these sources with more sensitive instruments for longer continuous integrations to determine flare timescales will allow us to perform better searches for fast X-ray variability in the future.  This will enable more complete studies of the connection between X-ray and TeV emission regions in blazar jets. \\

\begin{sidewaystable*}[!ht]
{%
\centering
\caption{The 16 potential quick time soft X-ray flares flagged for more analysis.\label{table:orbits}}
\begin{tabular}{c c c c c c c c c c c}
\hline\hline
Source & Date & Obs. Time & \multicolumn{1}{p{1.5cm}}{\centering Weighted \\ Mean \\ $\chi^2_{red}$} & \multicolumn{1}{p{1.5cm}}{\centering Quiescent \\ $\chi^2_{red}$} & T$_{double}$ & T$_{1/2}$ & \multicolumn{1}{p{1.5cm}}{\centering Weighted \\ Mean \\ p-value} & Post-Trials & \multicolumn{1}{p{1.5cm}}{\centering Quiescent \\  p-value} & Post-Trials \\ [0.3ex] 
\hline
\\																														
TXS 0954+658		& 04-Jul-06	& 53920.442	& 3.23 & 4.38 & -	& 9.1 & 3.96\e{-2} & 6.34\e{-1} & 1.25\e{-2} & 1.99\e{-1} \\
1ES 2344+514 	& 03-Jan-10	& 55199.419 & 4.26 & 6.70 & -	& 2.8 & 5.13\e{-3} & 8.20\e{-2} & 1.61\e{-4} & 2.58\e{-3} \\
3C 279 			& 31-Dec-13	& 56657.480 & 4.45 & 11.96& 3.4	& -	  & 3.95\e{-3} & 6.32\e{-2} & 7.91\e{-8} & 1.27\e{-6} \\
3C 279          & 10-Jul-06	& 53926.208	& 3.05 & 4.69 & -	& 3.9 & 2.73\e{-2} & 4.36\e{-1} & 2.81\e{-3} & 4.49\e{-2} \\
3C 345			& 23-Aug-09	& 55066.809 & 3.19 & 4.29 & -	& 7.0 & 4.10\e{-2} & 6.56\e{-1} & 1.37\e{-2} & 2.19\e{-1} \\
3C 454.3		& 02-Oct-13	& 56567.783 & 3.63 & 11.52& 4.2 & -   & 2.66\e{-2} & 4.26\e{-1} & 9.90\e{-6} & 1.58\e{-4} \\
PKS 0235+164 	& 06-Feb-07	& 54137.292	& 5.95 & 7.79 & 2.5 & -   & 2.60\e{-3} & 4.16\e{-2} & 4.13\e{-4} & 6.61\e{-3} \\
PKS 0537-441, 1 & 08-Oct-08	& 54747.866	& 4.37 & 6.10 & -	& 3.6 & 1.27\e{-2} & 2.03\e{-1} & 2.25\e{-3} & 3.60\e{-2} \\
PKS 0537-441, 2 & 08-Oct-08	& 54747.869	& 5.05 & 6.76 & 3.2 & -   & 6.41\e{-3} & 1.02\e{-1} & 1.16\e{-3} & 1.85\e{-2} \\
PKS 0537-441	& 26-Jan-05	& 53396.605	& 5.68 & 8.08 & 2.2 & -   & 3.40\e{-3} & 5.44\e{-2} & 3.11\e{-4} & 4.97\e{-3} \\
PKS 1424+240 	& 13-Apr-13	& 56395.315 & 3.50 & 5.45 & -	& 4.9 & 1.47\e{-2} & 2.35\e{-1} & 9.68\e{-4} & 1.55\e{-2} \\
PKS 1510-089 	& 28-Apr-11	& 55679.860 & 5.94 & 6.96 & -	& 5.8 & 2.63\e{-3} & 4.21\e{-2} & 9.46\e{-4} & 1.51\e{-2} \\
RX J0324.6+3410 & 15-Jan-13	& 56307.859 & 3.91 & 9.46 & 5.7 & -   & 3.57\e{-3} & 5.71\e{-2} & 1.21\e{-7} & 1.93\e{-6} \\
S5 0716+714     & 14-Dec-09	& 55179.704	& 4.05 & 6.62 & -	& 2.2 & 1.74\e{-2} & 2.78\e{-1} & 1.33\e{-3} & 2.13\e{-2} \\
S5 0716+714     & 15-Dec-09	& 55180.299	& 5.51 & 10.98& -	& 3.2 & 4.05\e{-3} & 6.48\e{-2} & 1.71\e{-5} & 2.73\e{-4} \\
W Com			& 09-Jun-10	& 55356.738	& 3.78 & 5.36 & -	& 8.1 & 2.28\e{-2} & 3.65\e{-1} & 4.68\e{-3} & 7.49\e{-2} \\
[1ex]
\hline 
\end{tabular}
}
\* Notes.---Col. (1): Object name referred to on \textit{Swift}-XRT monitoring site. Col. (2): Date of observation. Col. (3): Start time of observation in MJD. Col. (4): Reduced $\chi^2$ of weighted mean straight line fit. Col. (5): Reduced $\chi^2$ of simulated quiescent line fit. Col. (6): Estimated doubling time of potential flare in minutes. Col. (7): Estimated halving time of potential flare in minutes. Col. (8): Estimated p-value significance of weighted mean straight line fit before trials factor. Col. (9): Estimated p-value significance of weighted mean straight line fit after trials factor. Col. (10): Estimated p-value significance of quiescent line fit before trials factor. Col. (11): Estimated p-value significance of quiescent line fit after trials factor.
\end{sidewaystable*}

\end{document}